\documentclass[conference]{IEEEtran}

\usepackage{amsmath,graphicx}
\usepackage{amssymb}
\usepackage{algpseudocode}
\usepackage{cite}
\usepackage{amsfonts}
\usepackage{graphicx,subfigure}
\usepackage{fancyhdr}  %
\usepackage{cases}
\usepackage{extarrows}
\usepackage{algorithm}
\usepackage{multirow,tabularx}
\usepackage{mathtools}
\usepackage{xcolor}
\usepackage[english]{babel}

\IEEEoverridecommandlockouts   

\begin{document}

\title{Deep Learning for UL/DL Channel Calibration\\ in Generic Massive MIMO Systems}
\author{
\IEEEauthorblockN{Chongwen~Huang$^1$, George~C.~Alexandropoulos$^2$, Alessio Zappone$^3$, Chau Yuen$^1$, and M\'{e}rouane Debbah$^{3,4}$
\thanks{The work of Prof. C. Yuen was supported by the MIT-SUTD International design center and NSFC 61750110529 Grant, and that of C. Huang by the PHC Merlion PhD program. The work of Prof. M. Debbah was supported by H2020 MSCA IF BESMART, Grant 749336, and H2020-ERC PoC-CacheMire, Grant 727682; the former project also funded the work of A. Zappone. During the preparation of this work Prof. G. C. Alexandropoulos was with the Mathematical and Algorithmic Sciences Lab, Huawei Technologies France SASU, 92100 Boulogne-Billancourt, France.}
 }
\IEEEauthorblockA{
$^1$Singapore University of Technology and Design, 487372 Singapore\\
emails: chongwen\_huang@mymail.sutd.edu.sg, yuenchau@sutd.edu.sg\\
$^2$Department of Informatics and Telecommunications, National and Kapodistrian University of Athens, Greece\\
email: alexandg@di.uoa.gr\\
$^3$CentraleSup\'elec, Universit\'e  Paris-Saclay, 91192 Gif-sur-Yvette, France\\
email: alessio.zappone@unicas.it}
$^4$Mathematical and Algorithmic Sciences Lab, Huawei Technologies France SASU,92100 Boulogne-Billancourt,France\\
email: merouane.debbah@huawei.com
}

\maketitle


\begin{abstract}
One of the fundamental challenges to realize massive Multiple-Input Multiple-Output (MIMO) communications is the accurate acquisition of channel state information for a plurality of users at the base station. This is usually accomplished in the UpLink (UL) direction profiting from the time division duplexing mode. In practical base station transceivers, there exist inevitably nonlinear hardware components, like signal amplifiers and various analog filters, which complicates the calibration task. To deal with this challenge, we design a deep neural network for channel calibration between the UL and DownLink (DL) directions. During the initial training phase, the deep neural network is trained from both UL and DL channel measurements. We then leverage the trained deep neural network with the instantaneously estimated UL channel to calibrate the DL one, which is not observable during the UL transmission phase. Our numerical results confirm the merits of the proposed approach, and show that it can achieve performance comparable to conventional approaches, like the Agros method and methods based on least squares, that however assume  linear hardware behavior models. More importantly, considering generic nonlinear relationships between the UL and DL channels, it is demonstrated that our deep neural network approach exhibits robust performance, even when the number of training sequences is limited.
\end{abstract}



\section{Introduction}
Massive Multiple-input Multiple-Output (MIMO) is already considered as a core physical layer component for fifth Generation (5G), and beyond, broadband wireless networks \cite{5G,chongwenTSP2019,Larsson2014}. It refers to the deployment of very large number of antenna elements at the Base Station (BS), which are intended to simultaneously serve multiple User Equipments (UEs). As it has been shown, massive MIMO can drastically improve the spectral efficiency of cellular networks \cite{Rusek2013,Larsson2014,chognwenTVT}. To achieve its theoretical gains, this technology necessitates accurate knowledge of the channels between the BS and each user, in order for the former to realize optimum DownLink (DL) precoding. Leveraging the Time Division Duplexing (TDD) mode \cite{Rusek2013}, this knowledge can be efficiently acquired from the UpLink (UL) direction via orthogonal training signals simultaneously sent from the UEs to BS. The UL channels are first estimated, and then exploiting channel reciprocity \cite{Tse2002,ChongwenACCASP,Smith2004}, they are mapped to DL channels.

In practice, however, although in TDD mode the UL and DL wireless propagation channels are physically reciprocal \cite{Smith2004}, the analog front-end circuitry for transmission and reception at the BS and the mobile users is in general not \cite{calibration01,Robustcalibration01,ChongwenGlobcom,Petermann}. This complicates the application of the massive MIMO paradigm, rendering the effective baseband-to-baseband UL and DL channels non-reciprocal. To rely on the channel reciprocity assumption, and consequently, utilize the UL Channel State Information (CSI) estimation to compute the DL precoding vectors, the non-reciprocal transceiver responses need to be appropriately calibrated \cite{calibration02,calibration04,ChongwenTWC,calibration08,calibration07}. This process is often termed as channel calibration, and usually includes two procedural steps: \textit{i}) estimation of the calibration coefficients between the UL and DL channels; and \textit{ii}) calibration compensation by applying those coefficients to the UL channel estimates in order to obtain the estimation of the DL channels.

Channel calibration of TDD MIMO channels with small numbers of antennas has been a matter of study in recent years. Depending on the system setup and requirements, the approach adopted can take many forms. For example, \cite{calibration01} proposed a methodology based on bidirectional measurements between the two ends of a MIMO link to estimate suitable reciprocity calibration coefficients. This calibration approach falls into the class of Over-The-Air (OTA) calibration schemes, where users are involved in the calibration process. Recently, the authors in \cite{argos} working on the massive MIMO Argos prototype performed OTA calibration with the help of a reference antenna. The Argos calibration approach, however, is sensitive to the location of the reference antenna, and as a  consequence, it is not suitable for distributed massive MIMO scenarios. As an improved method, \cite{NPCmehtod} proposed a novel family of calibration schemes based on antenna grouping, which can greatly speed up the calibration process with respect to the classical approaches. In addition, some experimental data about the calibration coefficients were recently reported in \cite{calibration09,NPCmehtod}, giving an insight on how the impairments evolve in the time and frequency domains as well as with temperature. In the latter two papers, the hardware characteristics behind those coefficients were also discussed.

Deep learning methods have demonstrated significant improvements in various application fields in the last years. These methods are capable of outperforming human-level object detection in some tasks \cite{hekaiming}, and achieve state-of-the-art results in machine translation \cite{machinetrans} and speech processing \cite{sprnn}. Additionally, deep learning combined with reinforcement learning techniques was able to beat human champions in challenging games, such as the Go \cite{go}. Very recently,  deep learning methods have been also proposed for communication systems. For instance, various channel decoders using deep learning techniques were proposed in \cite{polar,decoded}. In \cite{autoencoder}, it was proposed to learn a channel auto-encoder via deep learning tools. However, the use of deep learning methods for channel calibration in MIMO systems has not been investigated, yet.

In this paper, we propose a deep learning framework for calibration between UL and DL channels in massive MIMO systems. The proposed framework is rather general, targeting at training a Deep Neural Network (DNN) to learn the, possibly nonlinear, relationship between UL and DL channels. To this end, our designed DNN can be applied to generic TDD and Frequency Division Duplexing (FDD) massive MIMO systems with nonlinear hardware transceivers. During the training phase, the DNN is trained from both UL and DL training measurements. Then, the trained network is used to carry out the calibration task for yet unobserved channels. Compared with existing methods \cite{argos, NPCmehtod} that are limited to linear UL/DL relationships, it is shown via simulations that our DNN design approaches the Cram\'{e}r-Rao Bound (CRB). Moreover our DNN approach exhibits robust performance, even when the number of training sequences is limited.

\section{System Model}
In this section, we introduce our considered generic massive MIMO system model together with the special case of linear channel model for TDD operation.

\subsection{Generic System Model}
In practical communication systems, the UL and DL baseband-to-baseband channels between any pair of nodes are usually nonlinear, due to the front-end hardware at Transmitter (TX) and Receiver (RX). For example, nonlinear solid-state devices, like signal power and low noise amplifiers, are modeled as memoryless devices whose nonlinearity appears when the instantaneous input signal power fluctuates and approaches the saturation level of the device \cite{Nonlinear}. Other nonlinear devices, such as fiber optics, are usually modeled as nonlinear functions with memory, and their nonlinear effects originate from the physical limitations of the communication channel \cite{Nonlinear01}.

\begin{figure}[!t]
\centering
\includegraphics[width=90mm]{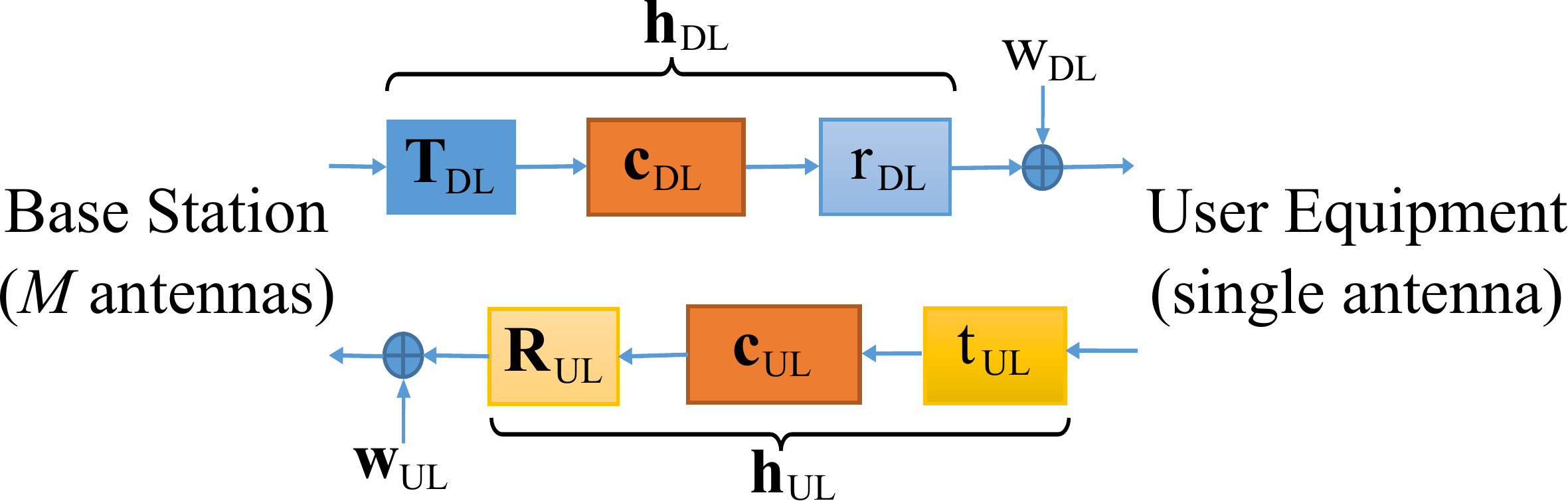}\\\vspace{-0.2cm}
    \caption{The considered model for bidirectional links (DL and UL directions) between a $M$-antenna base station and single-antenna user equipment.}
  \label{fig:model} \vspace{-0.2cm}
\end{figure}
Let us consider the bidirectional link depicted in Fig$.$~\ref{fig:model} between an $M$-antenna BS and a single-antenna User Equipment (UE). The upper part of the figure showcases the DL communication from BS to UE, whereas the lower part illustrates the UL communication from UE to BS. The $M\times M$ complex-valued matrices $\mathbf{T}_{\rm DL}$ and $\mathbf{R}_{\rm UL}$ represent the responses of the BS Radio Frequency (RF) front-ends of the transmit and receive modes, respectively. The diagonal elements in these matrices represent the linear effects attributable to the impairments in the TX and RX parts of the RF front-ends, respectively, whereas their off-diagonal elements correspond to RF crosstalk and possible antenna mutual coupling. The complex scalars $t_{\rm DL}$ and $r_{\rm UL}$ represent the responses of the UE RF front-end of the transmit and receive modes, respectively. Notations $\mathbf{c}_{\rm DL}\in\mathbb{C}^{1\times M}$ and $\mathbf{c}_{\rm UL}\in\mathbb{C}^{M\times 1}$ denote the OTA wireless propagation channels in the DL and UL directions, respectively. Without loss of generality, in this paper, we assume frequency flat channels, as typically occuring in narrowband communications or in a single subcarrier of a multi-carrier system. We have used $\mathbf{h}_{\rm DL} \in\mathbb{C}^{1\times M}$ and $\mathbf{h}_{\rm UL}\in\mathbb{C}^{M\times 1}$ to indicate the DL and UL propagation channels, respectively, in digital baseband. Finally, $\mathbf{w}_{\rm UL}\in\mathbb{C}^{M\times 1}$ denotes the Additive White Gaussian Noise (AWGN) vector at BS, while the complex scalar $w_{\rm DL}$ is the AWGN at the UE side.

Suppose that $N$ single UEs simultaneously transmit orthogonal pilot symbols in the UL to the $M$-antenna BS. By collecting all pilot symbols in the vector $\mathbf{x}_{\rm UL}\triangleq[x_{\rm UL}^{(1)}\,x_{\rm UL}^{(2)}\,\cdots\,x_{\rm UL}^{(N)}]^T$, where $x_{\rm UL}^{(n)}$ denotes the pilot symbol of the $n$-th UE with $n=1,2,\ldots,N$, the $M$-element complex-valued received signal at the BS antenna elements can be expressed in the general case as
\begin{equation}\label{RX_UL}
    \mathbf{y}_{\rm UL} = f(\mathbf{H}_{\rm UL}, \mathbf{x}_{\rm UL})+\mathbf{w}_{\rm UL},
 \end{equation}
where $\mathbf{H}_{\rm UL}\triangleq[\mathbf{h}_{\rm UL}^{(1)}\,\mathbf{h}_{\rm UL}^{(2)}\,\cdots\,\mathbf{h}_{\rm UL}^{(N)}]\in\mathbb{C}^{M\times N}$ (column concatenation) with $\mathbf{h}_{\rm UL}^{(n)}\triangleq[h_{\rm UL}^{(1,n)}\,h_{\rm UL}^{(2,n)}\,\cdots\,h_{\rm UL}^{(M,n)}]^T\in\mathbb{C}^{M\times 1}$ being the channel between BS and the $n$-th UE. Also, $f(\cdot)$ denotes a general nonlinear function of $\mathbf{H}_{\rm UL}$ and $\mathbf{x}_{\rm UL}$.

When DL channel estimation is the objective, the BS sequentially transmits the orthogonal pilot symbols vector $\mathbf{x}_{\rm DL}\triangleq[x_{\rm DL}^{(1)}\,x_{\rm DL}^{(2)}\,\cdots\,x_{\rm DL}^{(M)}]^T$ to each of the $N$ single-antenna UEs, with $x_{\rm DL}^{(m)}$ representing the pilot symbol transmitted from the $m$-th BS antenna. In this case, the $N$-element complex-valued vector including the received signals at all UE is given by the general expression
\begin{equation}\label{RX_DL}
    \mathbf{y}_{\rm DL}=g(\mathbf{H}_{\rm DL}, \mathbf{x}_{\rm DL})+\mathbf{w}_{\rm DL},  \\
 \end{equation}
where $g(\cdot)$ is a general nonlinear function of the DL baseband-to-baseband channel $\mathbf{H}_{\rm DL}\triangleq[\mathbf{h}_{\rm DL}^{(1)};\,\mathbf{h}_{\rm DL}^{(2)};\,\cdots\,;\mathbf{h}_{\rm DL}^{(N)}]\in\mathbb{C}^{N\times M}$ (row concatenation) and $\mathbf{x}_{\rm DL}$, and $\mathbf{w}_{\rm DL}\triangleq[w_{\rm DL}^{(1)}\,w_{\rm DL}^{(2)}\,\cdots\,w_{\rm DL}^{(N)}]\in\mathbb{C}^{N\times 1}$ represents the vector with the AWGN values at all $N$ UEs.

\subsection{Special Case: The Linear TDD Communication}\label{sec:Linear_Case}
State-of-the-art channel calibration methods \cite{calibration01, argos, NPCmehtod, calibration09} assume a linear relationship between the transmitted and received signals. Then, recalling our frequency flat channel model assumption, the received signals at the BS in the UL direction and at all UEs in the DL direction can be respectively expressed using \eqref{RX_UL} and \eqref{RX_DL} as
\begin{align} \label{mod_lin11}
\mathbf{y}_{\rm UL}&=\mathbf{H}_{\rm UL}\mathbf{x}_{\rm UL}+\mathbf{w}_{\rm UL}, \\
\mathbf{y}_{\rm DL}&=\mathbf{H}_{\rm DL}\mathbf{x}_{\rm DL}+\mathbf{w}_{\rm DL}.
\end{align}
By further assuming that our massive MIMO system operates in TDD mode, channel reciprocity holds for every link between BS and the $n$-th UE, i$.$e$.$, $\mathbf{c}_{\rm UL}^{(n)}=(\mathbf{c}_{\rm DL}^{(n)})^T$. In this case, the baseband DL and UL channels for every latter $n$-th link are respectively given according to Fig$.$~\ref{fig:model} as
\begin{align} \label{mod_lin22}
\mathbf{h}_{\rm DL}^{(n)}&=r_{\rm DL}^{(n)}\mathbf{c}_{\rm DL}^{(n)}\mathbf{T}_{\rm DL}^{(n)}, \\
\mathbf{h}_{\rm UL}^{(n)}&=\mathbf{R}_{\rm UL}^{(n)}(\mathbf{c}_{\rm DL}^{(n)})^T t_{\rm UL}^{(n)},
\end{align}
and hence, they can be easily related as follows:
\begin{align} \label{mod_lin33}
\mathbf{h}_{\rm DL}^{(n)}&=r_{\rm DL}^{(n)}\mathbf{c}_{\rm DL}^{(n)}\mathbf{T}_{\rm DL}^{(n)},  \nonumber \\
               &=r_{\rm DL}^{(n)}((\mathbf{R}_{\rm UL}^{(n)})^{-1}\mathbf{h}_{\rm UL}^{(n)}(t_{\rm UL}^{(n)})^{-1})^T\mathbf{T}_{\rm DL}^{(n)}, \nonumber \\
               &=\underbrace{r_{\rm DL}^{(n)}(t_{UL}^{(n)})^{-1}}_{\triangleq a^{(n)}}(\mathbf{h}_{\rm UL}^{(n)})^{T}
\underbrace{(\mathbf{R}_{\rm UL}^{(n)})^{-T}\mathbf{T}_{\rm DL}^{(n)}}_{\triangleq\mathbf{B}^{(n)}},  \nonumber \\
               &=a^{(n)}(\mathbf{h}_{\rm UL}^{(n)})^{T}\mathbf{B}^{(n)}.
\end{align}

The OTA channel reciprocity calibration for the case of massive MIMO TDD systems has the following two phases. In the first phase, the calibration process is performed which includes the estimations of the complex-valued scalar $a^{(n)}$ and $\mathbf{B}^{(n)} \in\mathbb{C}^{M\times M}$ appearing in \eqref{mod_lin33} $\forall$$n$. Then, during the second data transmission phase, the latter matrices are used together with the instantaneously measured UL channels $\mathbf{h}_{\rm UL}^{(n)}$ $\forall$$n$ to estimate $\mathbf{h}_{\rm DL}^{(n)}$ $\forall$$n$ according to \eqref{mod_lin33}.

\section{DNN-Based Channel Calibration}
In this section, we present a DNN-based channel calibration approach for generic massive MIMO systems, which is fully data driven. It comprises a training phase to obtain the designed DNN's parameters, followed by the online channel calibration phase that feeds UL channel estimation to the DNN, which finally outputs the predicted DL channels.

\subsection{Deep Learning Basics}
Suppose that $\boldsymbol{\theta}\triangleq\{\boldsymbol{\theta}_{1},\boldsymbol{\theta}_{2},\ldots,\boldsymbol{\theta}_{L}\}$ includes $L$ sets of parameters. A feedforward DNN (or multi-layer perceptron) with $L$ layers describes a mapping $F(\mathbf{r}_0,\boldsymbol{\theta}): \mathbb{R}^{N_0\times1}\mapsto \mathbb{R}^{N_L\times1}$ of the input vector $\mathbf{r}_0 \in \mathbb{R}^{N_0\times1} $ to an output vector in $\mathbb{R}^{N_L\times1}$ through the following $L$ iterative processing steps:
\begin{equation}\label{model_DL1}
  \mathbf{r}_{\ell}\triangleq f_{\ell}( \mathbf{r}_{\ell-1};\boldsymbol{\theta}_{\ell}),\,\,\ell=1,2,\ldots,L,
\end{equation}
where $f_{\ell}( \mathbf{r}_{\ell-1};\boldsymbol{\theta}_{\ell}): \mathbb{R}^{N_0\times1}\mapsto \mathbb{R}^{N_L\times1}$ represents the mapping carried out by the $\ell$-th DNN layer. This mapping depends on the output vector $\mathbf{r}_{\ell-1}$ from the previous $(\ell-1)$-th layer and on a set of parameters $\boldsymbol{\theta}_{\ell}$. In general, the mapping $f_{\ell}(\cdot;\cdot)$ can be stochastic, i$.$e$.$, it can be a function of random variables. The $\ell$-th DNN layer is called dense or fully-connected if all neurons in this layer are connected to all neurons in the following layer. In this case, $f_{\ell}( \mathbf{r}_{\ell-1};\boldsymbol{\theta}_{\ell})$ with $\boldsymbol{\theta}_{\ell}\triangleq\{\mathbf{W}_{\ell},\mathbf{b}_{\ell}\}$ has the form:
\begin{equation}\label{model_DL2}
f_{\ell}( \mathbf{r}_{\ell-1};\boldsymbol{\theta}_{\ell})=\sigma(\mathbf{W}_{\ell}\mathbf{r}_{\ell-1}+\mathbf{b}_{\ell}),
\end{equation}
where $\mathbf{W}_{\ell} \in \mathbb{R}^{N_{\ell}\times (N_{\ell}-1)}$ denotes the neurons' weights at this layer, $\mathbf{b}_{\ell} \in \mathbb{R}^{N_{\ell}\times1}$ stands for the bias vector, and $\sigma(\cdot)$ represents a so-called activation function.
\begin{figure}
\centering
\includegraphics[width=90mm]{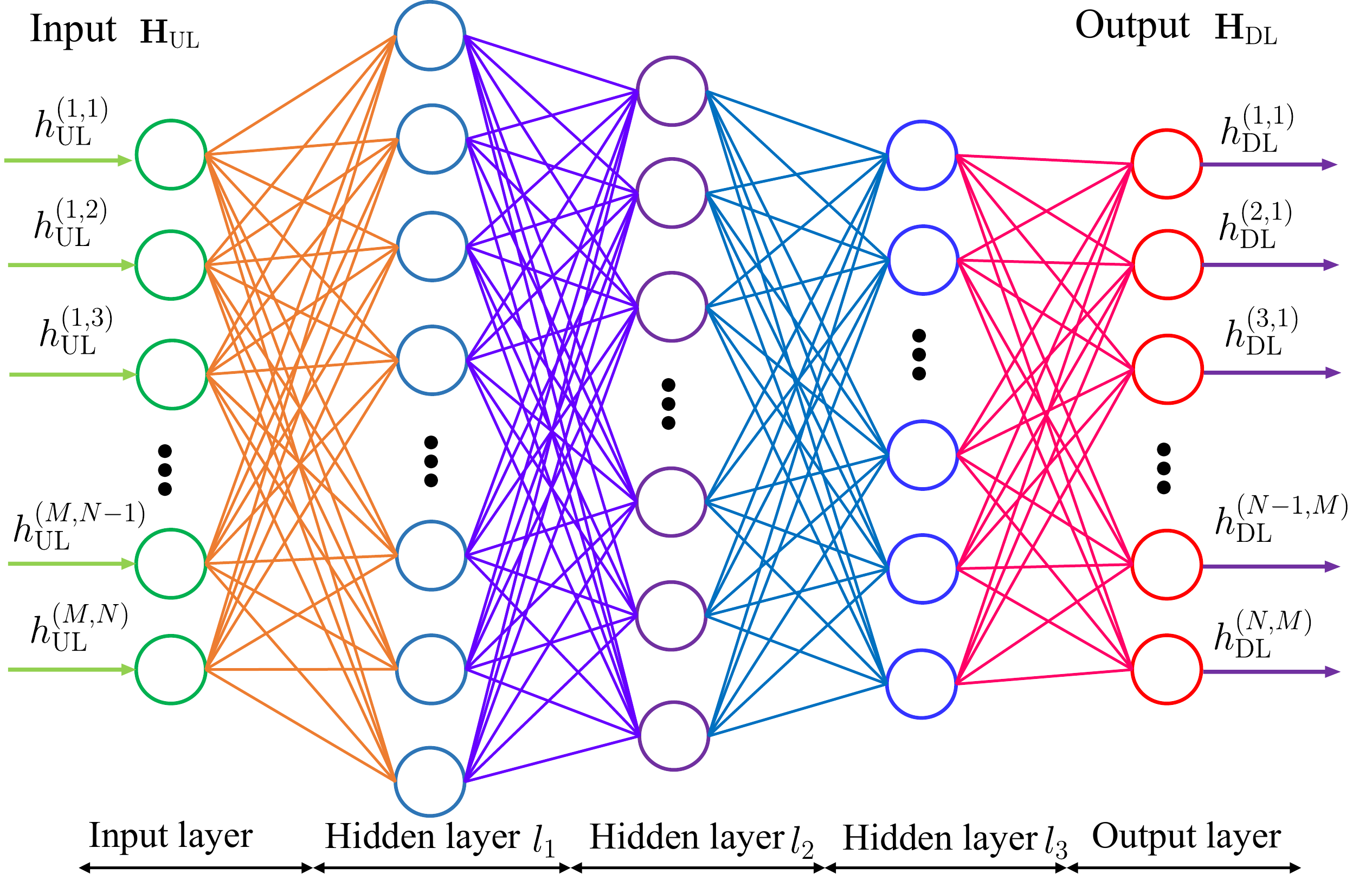} 
    \caption{The proposed Calinet DNN for channel calibration in generic massive MIMO systems comprising of five layers.}
    \label{fig:DNN} \vspace{-0.4cm}
\end{figure}

\subsection{Proposed DNN Massive MIMO Channel Calibration}\label{sec:proposed_DNN}
According to the \textit{universal approximation theorem} \cite{universal}, a feed-forward neural network with a fully-connected single hidden layer can approximate continuous functions on compact subsets of $\mathbb{R}^{N\times1}$. As a standard multi-layer processor, a DNN is capable of approximating any continuous function to any desired degree of accuracy. Our proposed DNN for channel calibration in generic massive MIMO systems, named \textbf{Calinet}, is illustrated in Fig$.$~\ref{fig:DNN}. It consists of five layers; namely, the input layer, three hidden layers, and the output layer. In particular, there exist multiple neurons in each hidden layer, and each layer's output is a nonlinear function of a weighted sum of the values of the neurons at the input of this layer. As shown in Fig$.$~\ref{fig:DNN}, Calinet's input layer consists of $MN$ neurons, which forward the instantaneous channel coefficients in $\mathbf{H}_{\rm UL}$ to the first hidden layer $l_1$. The role of the hidden and output layers is to capture the relationship between $\mathbf{H}_{\rm UL}$ and $\mathbf{H}_{\rm DL}$. In all layers, we have used $\mathrm{tanh}(\cdot)$ as activation function $\sigma(\cdot)$, since it works well with approximations of nonlinear functions and provides full mapping in $[-1,1]$. 
\begin{figure}
\centering
\includegraphics[width=85mm]{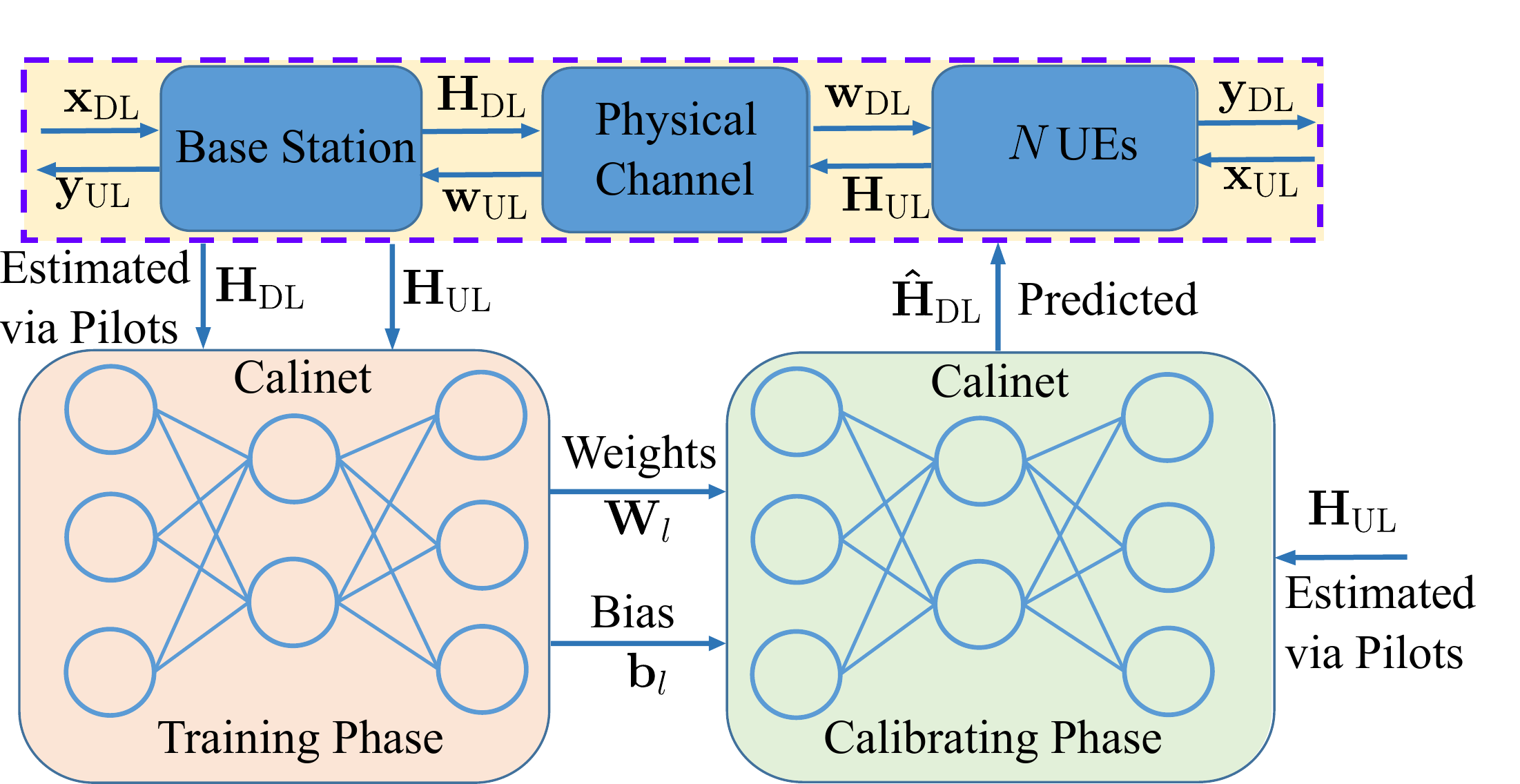}\\\vspace{-0.2cm}
    \caption{Schematic of the proposed procedure for deep learning based massive MIMO channel calibration using the Calinet DNN in Fig$.$~\ref{fig:DNN}.} \vspace{-0.2cm}
    \label{fig:flow}
\end{figure}

The proposed DNN-based massive MIMO channel calibration works as illustrated in Fig$.$~\ref{fig:flow}. First, the DNN of Fig$.$~\ref{fig:DNN} undergoes a training phase, and then, the trained DNN is used for online channel calibration. More specifically:
\begin{enumerate}
  \item \textit{Training Phase}: In this phase, pilot signals are used in both the DL and UL directions to obtain $P$ pairs of bidirectional channel estimations; let $(\mathbf{H}_{{\rm DL},p}, \mathbf{H}_{{\rm UL},p})$ denote the $p$-th pair, with $p=1,2,\ldots,P$, of UL and DL channel estimations. Then, this $P$-element training set is fed to Calinet and a training algorithm is deployed to adjust the DNN's parameters in $\boldsymbol{\theta}$. This process is detailed in the following subsection.
  \item \textit{Channel Calibration Phase}: The trained Calinet from the previous phase is used to predict DL channels from UL channel estimations. The UL channel estimation $\mathbf{H}_{\rm UL}$ obtained from UL pilots is fed to Calinet that outputs the predicted DL channels. We denote those channels by $\mathbf{\hat{H}}_{\rm DL}\triangleq[\mathbf{\hat{h}}_{\rm DL}^{(1)};\,\mathbf{\hat{h}}_{\rm DL}^{(2)};\,\cdots\,;\mathbf{\hat{h}}_{\rm DL}^{(N)}]\in\mathbb{C}^{N\times M}$.
\end{enumerate}

\subsection{DNN Training Process}\label{sec:training}
In the beginning of the training phase, the desired inputs $\{\mathbf{H}_{{\rm UL},p}\}_{p=1}^P$ and desired outputs $\{\mathbf{H}_{{\rm DL},p}\}_{p=1}^P$ of the DNN Calinet are obtained via dedicated channel sounding. Afterwards, the weights and bias of Calinet's layers are designed in order to minimize the Minimum Mean Squared Error (MMSE) between the $P$ actual outputs $\{\mathbf{\hat{H}}_{{\rm DL},p}\}_{p=1}^P$ of the DNN and the $P$ desired outputs $\{\mathbf{H}_{{\rm DL},p}\}_{p=1}^P$, namely
\begin{align}\label{model_DL3}
   & \mathcal{L}(\mathbf{\hat{H}}_{{\rm DL},p}(\boldsymbol{\theta}),\mathbf{H}_{{\rm DL},p}) = \notag\\
   &=\sum_{p=1}^{P}\|vec(\mathbf{\hat{H}}_{{\rm DL},p}(\boldsymbol{\theta}))-vec(\mathbf{H}_{{\rm DL},p})\|^2
\end{align}
where $\boldsymbol{\theta}\triangleq\{\boldsymbol{\theta_{1}},\boldsymbol{\theta_{2}},\ldots,\boldsymbol{\theta_{5}}\}$ with $\boldsymbol{\theta_{\ell}}=\{\mathbf{W}_{\ell},\mathbf{b}_{\ell}\}$. Notation $\mathbf{\hat{H}}_{{\rm DL},p}(\boldsymbol{\theta})$ indicates that the actual DNN vector output is a function of the network's parameters $\boldsymbol{\theta}$ and of course its input vector $\mathbf{H}_{{\rm UL},p}$.

In order to minimize \eqref{model_DL3} with respect to $\boldsymbol{\theta}$, several off-the-shelf training algorithms can be used, that employ variations of the stochastic gradient descent methods, where at each step the gradient of the cost function is estimated from a randomly-selected subset of training samples, called mini-batch. Thus, at each gradient iteration, the DNN parameters are updated as
\begin{align}\label{training12}
 \boldsymbol{\theta}_i = \boldsymbol{\theta}_i -\eta \widehat{\nabla} \mathcal{L}(\boldsymbol{\theta}_{i-1}),
\end{align}
where $\widehat{\nabla} \mathcal{L}(\boldsymbol{\theta}_{i-1})$ denotes the estimated gradient, and $\eta$ the algorithm learning rate. Moreover, at each step, gradients are efficiently computed by means of the  backpropagation method \cite{backpropagation}. Specifically, the proposed DNN Calinet has been trained by means of the Adagrad algorithm \cite{adagrad}. 

\begin{table}[!t]
\caption{Parameters for Calinet Design and Simulations} \label{tabpar} \vspace{-4mm}
\begin{center}
    \begin{tabular}{| l | l |}
    \hline
    \textbf{Parameters} & \textbf{Values} \\ \hline\hline
     Input layer dimension  & 32 \\ \hline
     Learning rate & 0.01 \\ \hline
     Optimization algorithm & Adagrad \cite{adagrad} \\ \hline
     Number of data damples  & 10240 \\ \hline
     Epoch & 256 \\ \hline
     Activation function & $\mathrm{tanh}(\cdot)$ \\ \hline
		 $r_{\rm DL}$, $t_{\rm UL}$ for all UEs & $\mathcal{CN}(0,1)$  \\ \hline
		 $\mathbf{R}_{\rm UL}$, $\mathbf{T}_{\rm DL}$ at BS & All elements in $\mathcal{CN}(0,1)$\\ \hline
     Validation split ratios & 0.4 \\ \hline
     Batch size & 4 \\ \hline
     SNR during training & 0dB-40dB \\ \hline
		$\mathbf{H}_{\rm UL}$, $\mathbf{H}_{\rm DL}$ & All elements in $\mathcal{CN}(0,1)$\\ \hline
     AWGN Distribution & $\mathcal{CN}(0,1/SNR)$ \\ \hline
     Pilot symbols $\mathbf{x}_{\textrm{DL}}$ and $\mathbf{x}_{\textrm{UL}}$ & Unit magnitudes, phases in $[-\pi,\pi]$ \\
    \hline\hline
    \end{tabular}
\end{center} \vspace{-4mm}
\end{table}

\section{Numerical Results}
For our numerical analysis, we consider a massive MIMO system comprising a BS with $M=32$ antenna elements and $N=4$ single-antenna UEs. Considering the linear TDD case of Sec$.$~\ref{sec:Linear_Case}, we compare the performance of the Argos \cite{argos} method and the Norm plus Phase Constraint First Coefficient (NPC FC-II) \cite{NPCmehtod} method, as well as the method in \cite{calibration04} with that of the proposed DNN approach presented in Secs$.$~\ref{sec:proposed_DNN} and~\ref{sec:training}. Moreover, we also report the CRB, which can be obtained in closed form following similar steps as in \cite{NPCmehtod}. We have also investigated the calibration performance of the proposed DNN approach for generic nonlinear scenarios, for which no other method is available. The simulation parameters and the parameters for the designed Calinet are summarized in Table~\ref{tabpar}, where SNR denotes the Signal-to-Noise Ratio and $\mathcal{CN}(0,\lambda)$ represents the circularly symmetric complex Gaussian distribution with zero mean and variance $\lambda$. A total of $500$ independent channel realizations were generated for the results that follow, which were obtained using the deep learning library TensorFlow \cite{tensorflow} and MATLAB$^{\rm TM}$. For the DNN approach, the considered data set was randomly partitioned into the training set (60\% of the total samples) and the test set (40\% of the total samples).
\begin{figure}[!t]
\centering
\includegraphics[width=88mm]{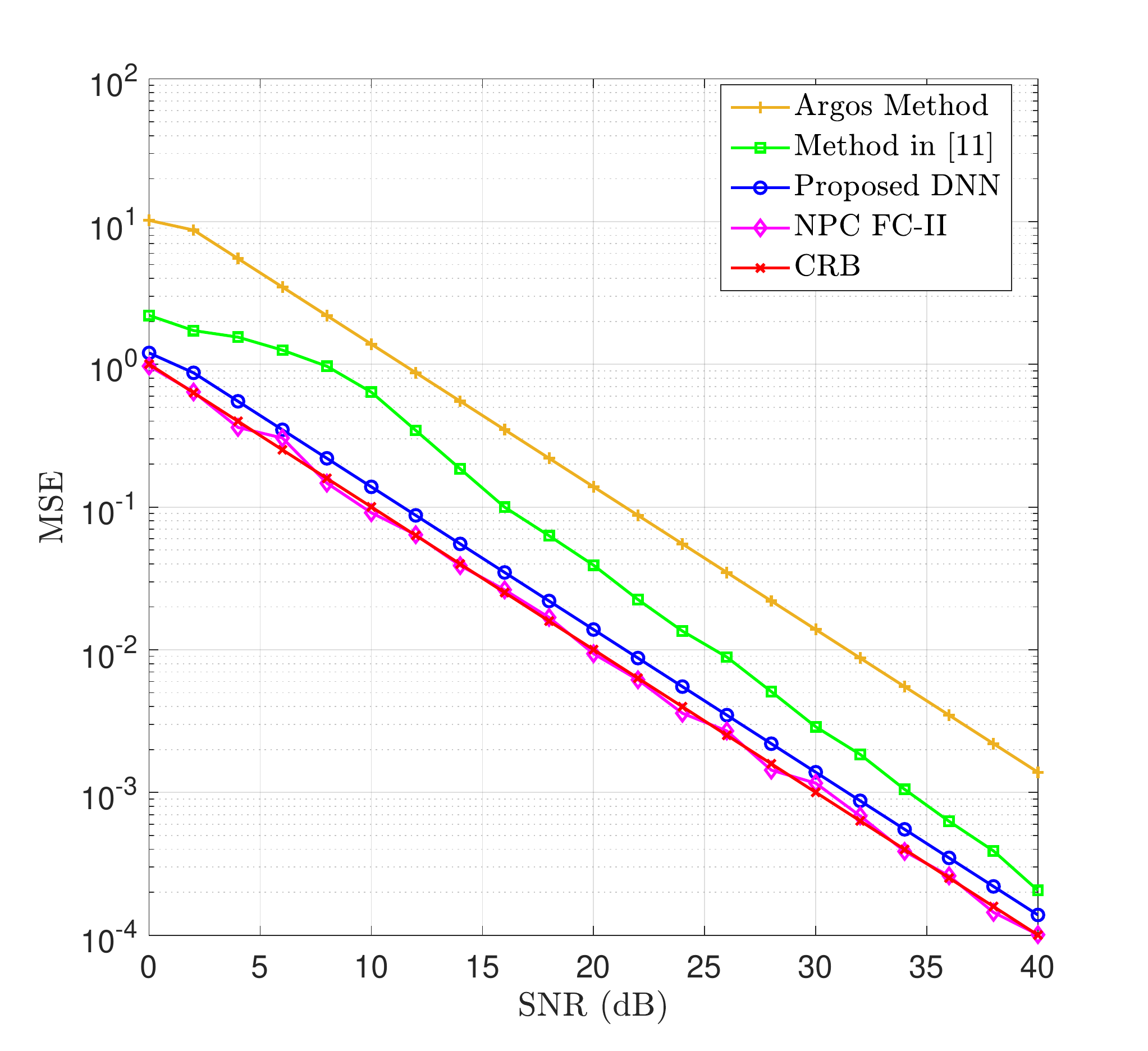}
    \caption{MSE comparison of various channel calibration schemes for a linear TDD massive MIMO system with $M=32$ and $N=4$.}
    \label{fig:comp} \vspace{-0.5cm}
\end{figure}

\begin{figure}[!t]
\centering
\includegraphics[width=88mm]{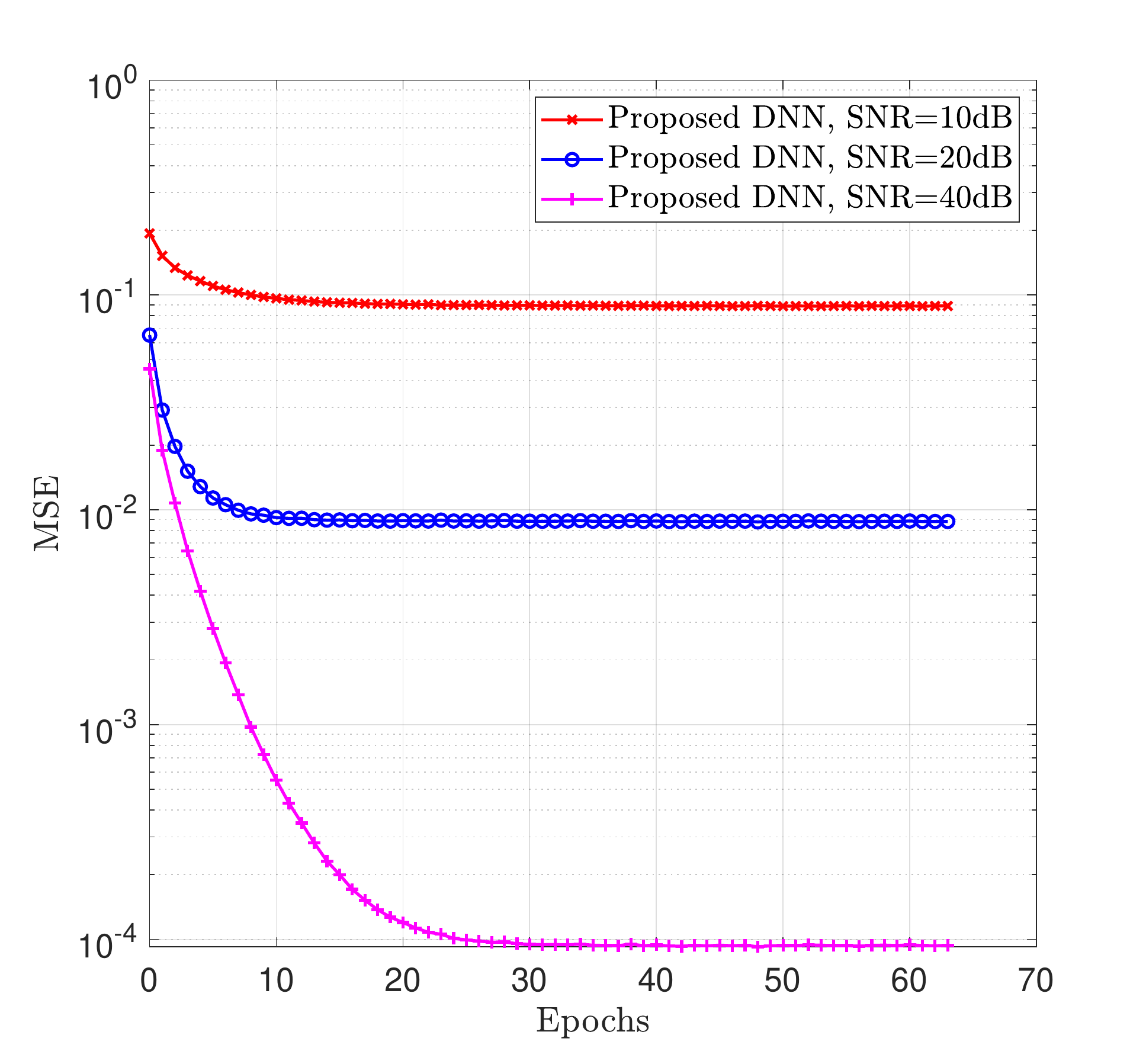}
    \caption{MSE performance convergence of the training process with the proposed DNN approach in Fig$.$~\ref{fig:comp} for different SNR values.}
    \label{fig:SNR} \vspace{-0.3cm}
\end{figure}

In Fig$.$~\ref{fig:comp}, the Mean Squared Error (MSE) performance between the actual $\mathbf{H}_{\rm DL}$ and the $\widehat{\mathbf{H}}_{\rm DL}$ obtained using all considered calibration methods is plotted as a function of the SNR in dB. As shown, all methods share the same decreasing MSE trend when SNR increases. Although the NPC FC-II algorithm achieves the best MSE performance since it is sufficiently close to CRB, the proposed DNN approach outperforms the Argos method and the approach from \cite{calibration04}, exhibiting only a small gap from CRB, despite the limited size of the training set. In Fig$.$\ref{fig:SNR}, we sketch the iteration behavior of the DNN training process for three different SNR values. It can be seen that MSE decreases drastically with more iterations and reaches a floor that becomes lower as the SNR increases. In addition, the lower the SNR the faster is the convergence. These results showcase that higher operating SNR values grant better DNN-based performance, but this happens at the price of a longer training time.

\begin{figure}[!t]
\centering
\includegraphics[width=88mm]{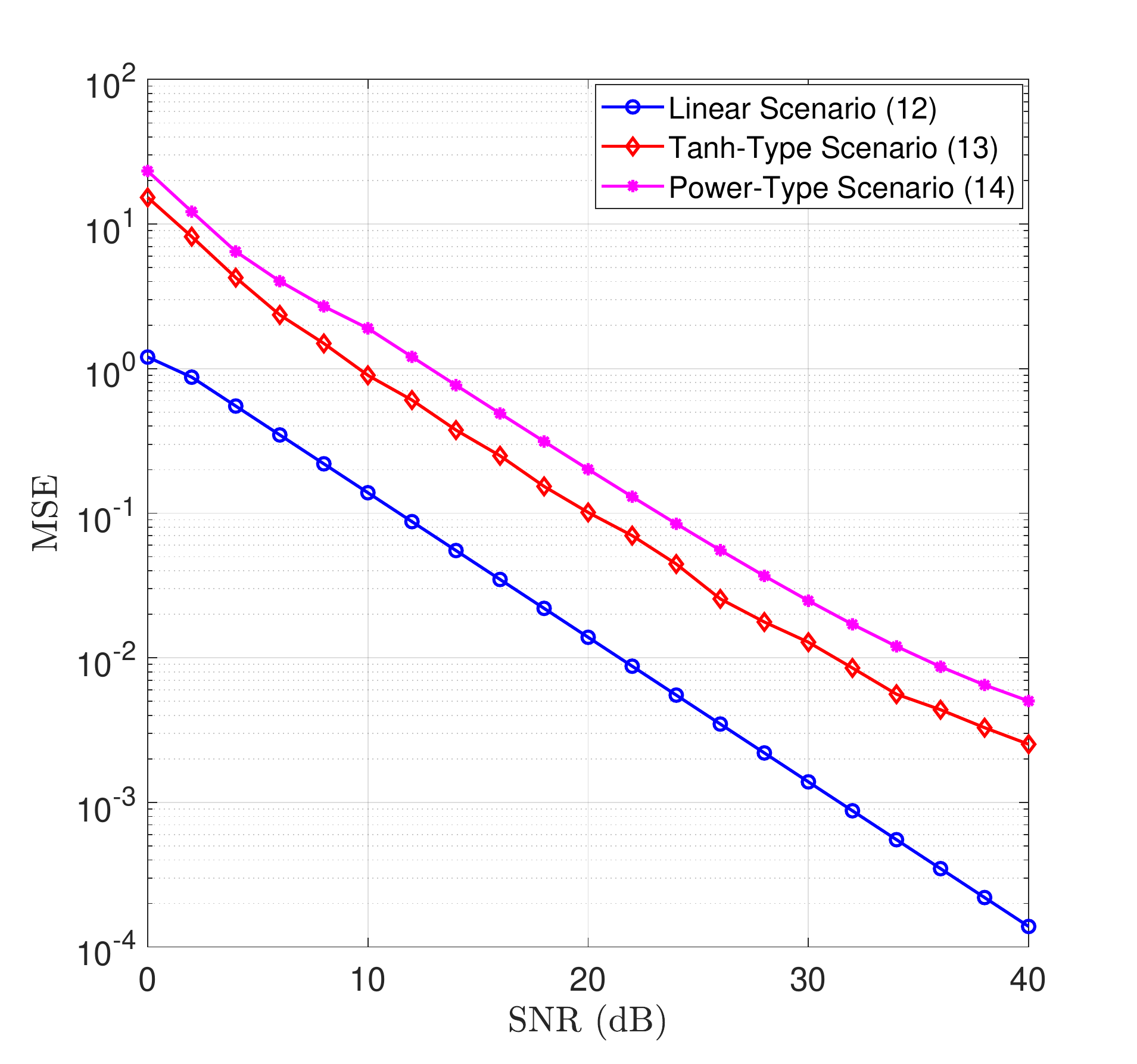}
    \caption{MSE performance of the proposed DNN-based channel calibration approach for generic massive MIMO scenarios with $M=32$ and $N=4$.}
    \label{fig:nonlinear} \vspace{-0.3cm}
\end{figure}

The MSE performance versus the SNR in dB of the proposed DNN-based channel calibration approach for generic massive MIMO scenarios is depicted in Fig$.$~\ref{fig:nonlinear}. We have assumed that the DL and UL channels between BS and each $n$-th UE exhibit the one linear (for TDD mode) and the two nonlinear relationships that follow:
\begin{equation} \label{non_simula1}
\mathbf{h}_{\rm DL}^{(n)}=c^{(n)}(\mathbf{h}_{\rm UL}^{(n)})^{T}\mathbf{D},
\end{equation}
\begin{align}  \vspace{-0.3cm}
\mathbf{h}_{\rm DL}^{(n)}&=c^{(n)}\mathrm{tanh}(\mathbf{h}_{\rm UL}^{(n)})\mathbf{D}, \label{eq:tanh-type} \\
\mathbf{h}_{\rm DL}^{(n)}&=c^{(n)}(\mathbf{h}_{\rm UL}^{(n)})^2\mathbf{D}, \label{eq:power-type}
\end{align}
where $c^{(n)}$ is a complex scalar distributed as $\mathcal{CN}(0,1)$ and $\mathbf{D}\in\mathbb{C}^{M\times M} $ is a unitary matrix. As shown in this figure, the MSE performance improves as the SNR increases for all considered linear and nonlinear scenarios. Moreover, it can be seen that the prediction error is larger for the nonlinear scenarios, especially the Power-Type one given by \eqref{eq:power-type}.

\begin{figure}[!t]
\centering
\includegraphics[width=88mm]{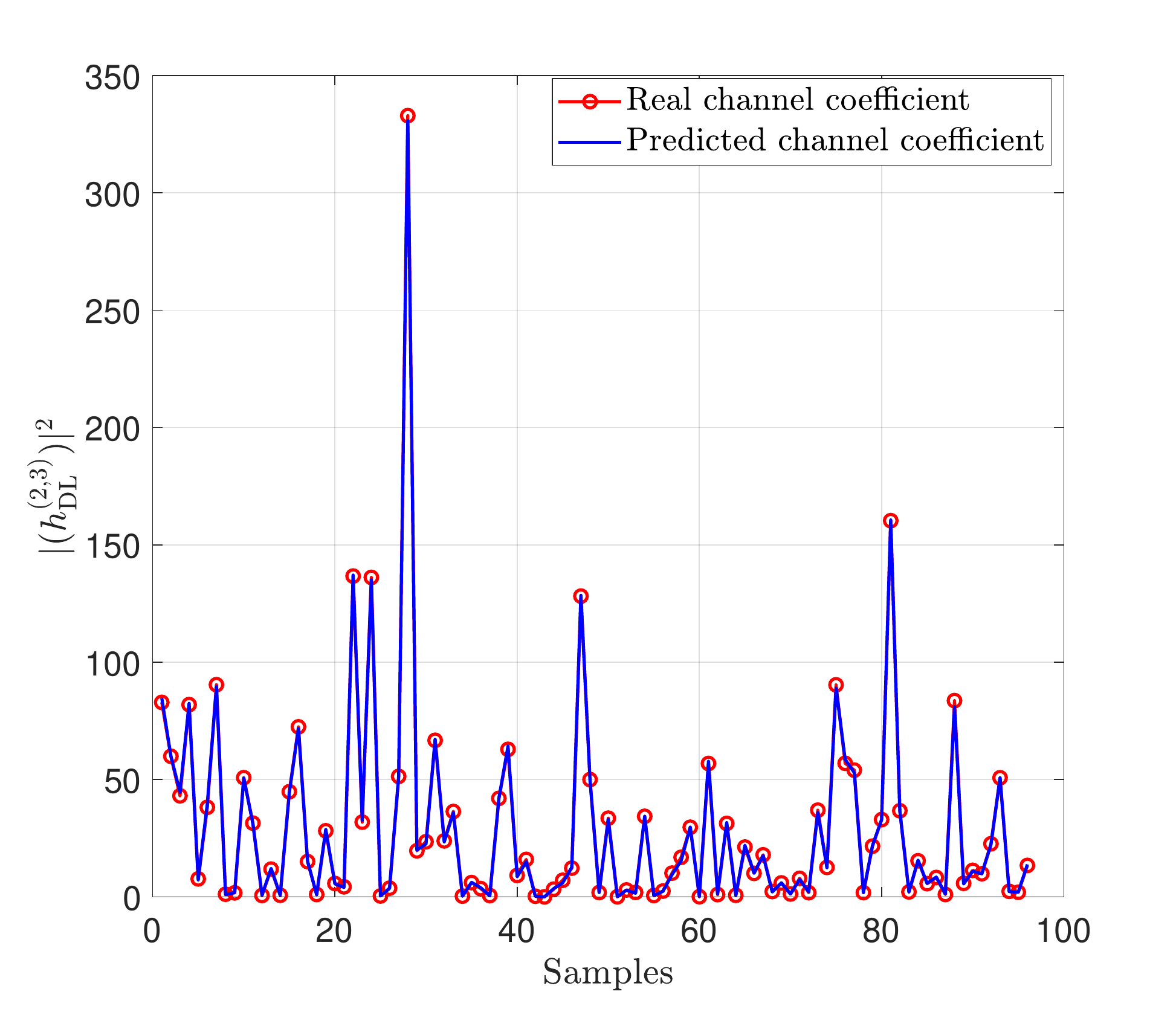}
    \caption{Predictions of a DL channel coefficient using the proposed DNN approach for the nonlinear Tanh-Type massive MIMO scenario \eqref{eq:tanh-type} of Fig$.$~\ref{fig:nonlinear} with operating SNR$=20$dB.}
    \label{fig:amp} \vspace{-0.3cm}
\end{figure}

Finally, Fig$.$~\ref{fig:amp} compares the actual and predicted values of the squared moduli of the channel coefficient $h_{\rm DL}^{(2,3)}$ between the $2$-nd BS antenna and the $3$-rd user's single-antenna. The nonlinear Tanh-Type massive MIMO scenario \eqref{eq:tanh-type} of Fig$.$~\ref{fig:nonlinear} was considered and the operating SNR was set to $20$dB. As shown in the figure, all predicted channel values sufficiently match the actual ones.

\section{Conclusion}
A deep-learning-based method for channel calibration in generic massive MIMO systems is proposed and compared against traditional calibration methods. Unlike existing methods that assume linear relationships between DL and UL channels, the proposed DNN method can operate also in nonlinear settings, performing close to the Cram\'{e}r-Rao bound in the linear scenario. More importantly, it exhibits robustness in generic nonlinear scenarios even when the number of training sequences is limited. Our results indicate that deep learning-based methods can have significant potential in many parameter estimation problems for communications, e.g., nonlinear channel estimation in massive MIMO systems.

\end{document}